\documentstyle{mn}
\newcommand{\sige}{{\mbox{$\sigma_{\!{\rm\tiny RA}}$}}}
\newcommand{\sigh}{{\mbox{$\sigma_{\!{\rm\tiny Dec}}$}}}
\newcommand{\rmoon}{{\mbox{$r_{\rm\tiny Moon}$}}}
\newcommand{\hr}{$^{\rm h}$}
\newcommand{\mn}{$^{\rm m}$}
\newcommand{\rs}{$^{\rm s}$}
\newcommand{\dg}{$^\circ$ }

\newcommand{\uk}{{\rm $\mu$K\,} }

\newcommand{\DT}{$\Delta T $\,}
\def\et{et~al.~}

\title{A measurement at the first acoustic peak of the CMB with the 33 GHz interferometer}
\author[Harrison et al.]{ D.L. Harrison$^1$, J. A. Rubi\~no-Martin$^2$,
S.J. Melhuish$^1$, R.A. Watson$^1$,\newauthor R.D. Davies$^1$, R. Rebolo$^{2,3}$,
R.J. Davis$^1$, C. M. Guti\'errez$^2$, J. F. Macias-Perez$^1$ 
\\
$^1$University of Manchester, Jodrell Bank Observatory, Macclesfield, Cheshire SK11 9DL, UK\\
$^2$Instituto de Astrofisica de Canarias, 38200 La Laguna, Tenerife,
Canary Islands, Spain \\
$^3$Consejo Superior de Investigaciones Cientificas, Spain.\\
}
\begin{document}
\maketitle
\begin{abstract}
	This paper presents the results from the Jodrell Bank~--~IAC
	two-element 33~GHz interferometer operated with an element
	separation of 32.9 wavelengths and hence sensitive to 1\dg-scale
	structure on the sky. The level of CMB fluctuations, assuming a
	flat CMB spatial power spectrum over the range of multipoles $
	\ell = 208 \pm 18 $, was found using a likelihood analysis to be
	$\Delta T_\ell=63^{+7}_{-6}{\rm\,\mu K}$ at the 68\%\,
	confidence limit, after the subtraction of the contribution of
	monitored point sources. Other possible foreground contributions
	have been assessed and are expected to have negligible impact on
	this result.

\end{abstract}

\begin{keywords}
cosmology: cosmic microwave background -- cosmology: observations --
large-scale structure of the Universe -- instrumentation: interferometers.
\end{keywords}

\section{Introduction}
\label{intro} 

Observations of the angular power spectrum of the CMB temperature
fluctuations are a powerful probe of the fundamental parameters of our
universe. The amplitude and spatial distribution of these fluctuations
can discriminate between competing cosmological models. Most
inflationary models predict more power on scales of 0\fdg2~--~2$^\circ$,
in the form of a series of peaks. These are due to acoustic
oscillations in the photon-baryon fluid, which are frozen into the CMB
at recombination, with the peaks corresponding to regions of maximum
compression and troughs regions of maximum rarefaction. Hence, the
position of the first acoustic peak is a strong test for the geometry of
the universe, since it corresponds to a fixed physical scale at the time of recombination projected onto the sky.

The previous result from the Jodrell~Bank~--~IAC 33~GHz interferometer
of $\Delta T_\ell=43^{+13}_{-12}{\rm\,\mu K}$, reported in Dicker
\et (1999), corresponds to an angular spherical harmonic $\ell \sim
110$, equivalent to $\sim$ 2\dg structure. To investigate smaller angular scales the baseline was doubled; in
this paper we analyze the data from this wide spacing configuration
which corresponds to an angular spherical harmonic $\ell \sim 210$. The data presented here were
taken at the Teide Observatory, Tenerife, between 27 May 1998 and 9
March 1999. The paper is organised as follows. The instrumental
configuration is summarised in Section~\ref{int33}; a full description
can be found in Melhuish \et (1999). The basic data processing is
outlined in Section~\ref{bdpcal}; for a more complete discussion see Dicker
\et (1999). The calibration method is also discussed in
Section~\ref{bdpcal} and the data analysis in Section~\ref{like}. A
derivation of the fluctuation amplitude, after an estimate of the
contribution of possible foregrounds, is given in Section~\ref{fore}.

\section{The 33~GHz Interferometer}
\label{int33}

\begin{figure*}
\begin{center}
\setlength{\unitlength}{1cm}
\begin{picture}(20,10.5)(0,0)
\put(0.7,-6){\includegraphics{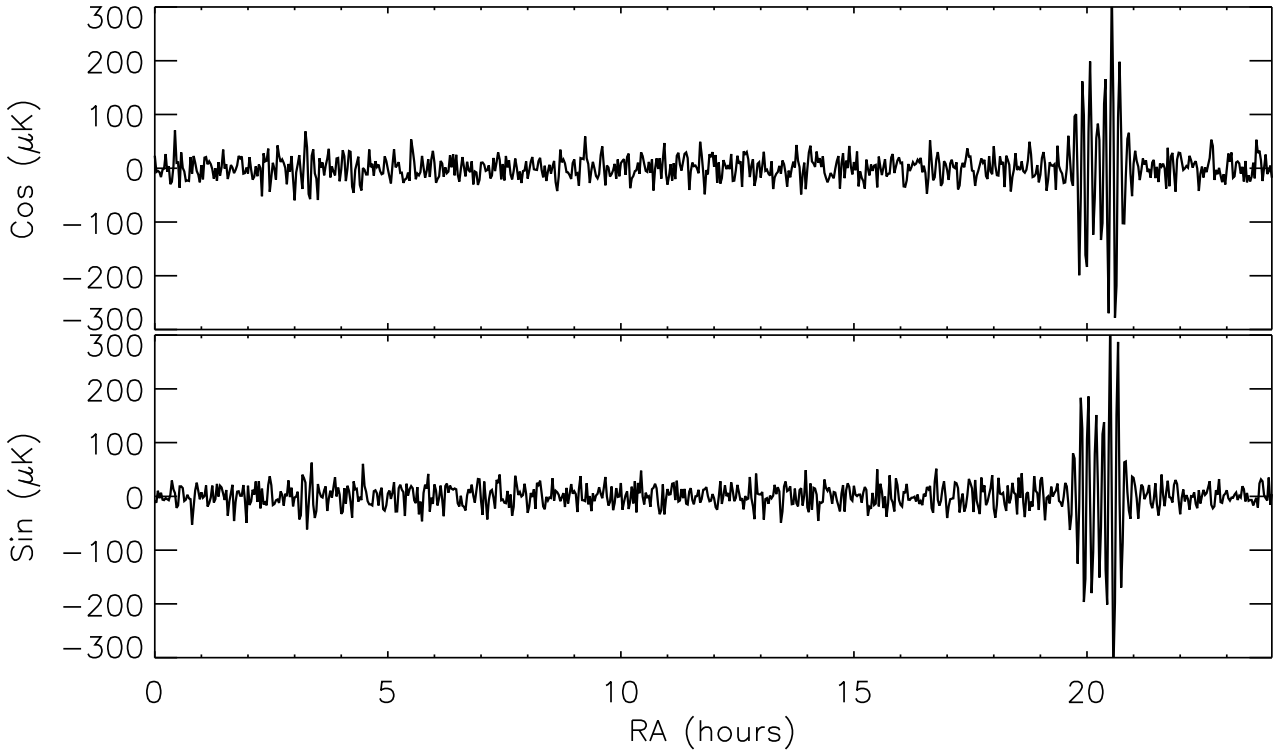}}
\end{picture}
\end{center}
\caption{The total data stack, in 0.5\dg-bins, at Dec +41\dg collected
over the period 27 May 1998 to 9 March 1999 and showing the cosine and
sine visibility data. The number of days' observations included at each
RA is between 180~--~210 days due to the removal of the Sun during the year. The Galactic plane crossing at the anti-centre is clearly visible at $21^{\rm{h}}$ together with Cyg A at $20^{\rm{h}} 30^{\rm{m}}$ and 3C84 at $3^{\rm{h}} 20^{\rm{m}}$}
\label{stack}
\end{figure*}

The interferometer consists of two  horn-reflector antennas  positioned
to form a single E--W baseline, which has two possible lengths depending
on the separation of the horns. The narrow spacing configuration has a
baseline of 152\,mm while in its wide spacing configuration the horns
are 304\,mm apart. For the observations presented here, the baseline was
304\,mm. Observations were made at a fixed declination of
Dec=+41\dg, using the rotation of the Earth
to ``scan'' 24\hr\ in RA each day. This ``scan'' runs through some of the
lowest background levels of synchrotron, dust and free-free emission. The horn
polarization is horizontal -- parallel with the scan direction. There
are two data outputs representing the cosine and the sine components of the
complex interferometer visibility. The operating bandwidth covers 
31\,~--~\,34\,GHz,  near a local minimum in the atmospheric emission spectrum; the
antenna spacing corresponds to 32.9 wavelengths. The low level of
precipitable water vapour, which is typically around 3\,mm at Teide
Observatory, permits the collection of high quality data. Only 16 per
cent of the data have been rejected, due to bad weather and the daily Sun transit.

The measured response of the interferometer is well approximated by a
Gaussian with sigmas of \sige = 2\fdg25 $\pm$ 0\fdg03 (in RA) and
\sigh = 1\fdg00 $ \pm $ 0\fdg02 (in Dec),  modulated by fringes with a period of $f$ = 1\fdg74 $ \pm $ 0\fdg02 in RA. This defines the range of sensitivity to the different multipoles $\ell$ of the CMB power spectrum ($C_{\ell}$) in the range corresponding to a maximum sensitivity at $\ell$ = 208 (0\fdg8) and half sensitivity at $\Delta \ell=\pm$18.

A known calibration signal (CAL) is periodically injected into the waveguide after the horns allowing a continuous calibration and concomitant corrections for drifts in the system gain and phase offset.

\section{Basic Data Processing and Calibration}
\label{bdpcal}

The first step in the analysis is the removal of any variable baseline
offsets from the data and the correction of a small departure from
quadrature between the cosine and sine data. The data are calibrated
relative to the CAL signal and re-binned into 2~-~minute bins to ensure
alignment in RA between successive scans. The data affected by the Sun
and bad weather are removed and individual scans are weighted, with
respect to their RMS error, to form a ``stack''. The total stack of all
the data used for this analysis is shown in Fig.~ \ref{stack}. The
number of observations at each RA varies from 210 to 180 days due the
removal of $ \pm 0.5^{h} $ about the Sun transit each day. 

The data are calibrated relative to CAL, although CAL itself needs to be calibrated
by an astronomical source. The large size of the primary beam results in
a reduced sensitivity to point sources and many days of observation are required to achieve a signal~-~to~-~noise ratio sufficient for calibration purposes. Consequently, the Moon is used as the primary calibrator as the power received from a single Moon transit is large enough to give signal~-~to~-~noise ratios of $ \sim 6000 $.\\

The Moon was modelled as a uniform disk of radius $\rmoon$\ and a 33~GHz
brightness temperature, $T_{\rm b}$ given by:

\begin{equation}
T_{\rm b}=202 + 27 \cos (\phi - \epsilon) \:\:\mbox{K}
\end{equation}where $\phi$\ is the phase of the Moon (measured from full Moon) and
$\epsilon$ = 41\dg\ is a phase offset caused by the finite thermal
conductivity of the Moon \cite{Gorenstein81}. It is sufficient to model
the Moon as a uniform disk when correcting for its partial
resolution by the interferometer. The effect of temperature variations across the
Moon are negligible when compared to the error, 5.5 per cent,  of the
Gorenstein and Smoot model for the Moon's brightness temperature. The expected antenna
temperature, $T_{\rm E}$, can then be found by integrating over the disk of
the Moon, multiplied by the normalized interferometer beam function:

\begin{eqnarray} 
\label{dilute} 
T_{\rm E} & = & \frac{T_{\rm b}}{2\pi\sige\sigh} \\ \nonumber
& \times & \int_0^{2\pi}\!\!\!\!\!\int_0^\rmoon \!\!\!\!\!\!\!\!\!\! \exp\left(
{-\frac{x^2}{2\sige^2}} {-\frac{y^2}{2\sigh^2}}\right)\; \cos \left(\frac{2\pi
x}{f}\right) \; {\rm d} \Omega 
\end{eqnarray}where $x = r\cos\psi$\ and $y = r\sin\psi$.  \sige\ and
\sigh\ are the RA and Dec beam sigmas (dispersion) and $f$ is the fringe spacing.

\begin{figure}
\begin{center}
\setlength{\unitlength}{1cm}
\begin{picture}(5,6)(0,0)
\put(-1.3,-2){\includegraphics{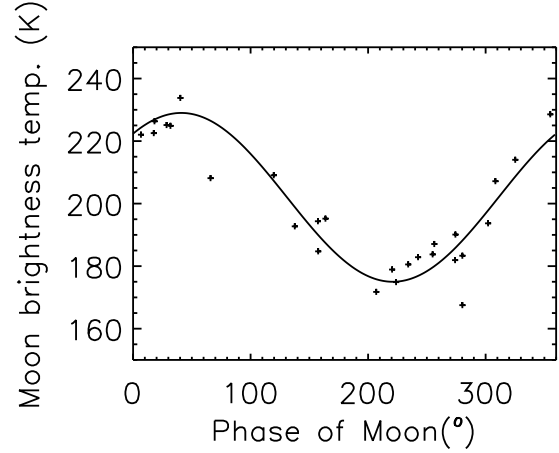}}
\end{picture}
\end{center}
\caption{Measured value of the Moon's brightness temperature as a function of phase. Phase is measured from the full Moon. The solid line is the predication of the model given by Gorenstein \& Smoot (1981): $ T_{\rm b}=202 + 27 \cos (\phi - \epsilon) $ in Kelvin. Each observation has been calibrated using CAL = 14.7 K.}
\label{moon1}
\end{figure}

Regular observations of the Moon were made; for each observation,
equation~\ref{dilute} was evaluated numerically and an amplitude for
CAL found such that the amplitude of the Moon in the processed data was
equal to the predicted value. Using 27 observations of the Moon, an
average amplitude for CAL of 14.7 $\pm$ 0.8\,K was found.  The error consists of  a 1.4 per cent error in the measurements and an estimated
5.5 per cent in the moon model \cite{Gorenstein81}. Fig.~\ref{moon1}
shows our observations and how the measured brightness temperatures of
the Moon change with phase. 

The moon model used in this paper differs
from the model given in Dicker~\et (1999). This data together with the addition
data taken in the narrow spacing, section~\ref{conclusions}, will be the
subject of a forthcoming paper, using the Moon model given in this paper.

\section{Likelihood Analysis}
\label{like}

\subsection{Theory}
\label{theory}

The temperature anisotropies of the CMB fluctuations are described by a
two~dimensional random field on the sky, the properties of which can be
determined from the two~point correlation function $C^{\rm
CMB}(\theta_{ij})$, where $\theta_{ij}$ is the angular separation of the two points.

\begin{equation}
 C^{\rm CMB} \left( \theta _{ij} \right) \equiv {\Big \langle}
 \left(\frac{\Delta T_{i}}{T}\right) \left(\frac{\Delta T_{j}}{T}\right)
 {\Big \rangle}
\end{equation} which can be expanded in terms of spherical harmonics as:

\begin{equation}
C^{\rm CMB}(\theta_{ij})=\sum^{\infty}_{2} \frac{(2\ell+1)}{4\pi} C_{\ell} P_{\ell}(\cos(\theta_{ij}))
\end{equation} In our analysis the form of the likelihood function is given by :
\begin{equation}
\label{eqn_like}
L \propto \frac{1}{{\bracevert C \bracevert}^{\frac{1}{2}}} {\rm exp} \left( - \frac{1}{2}D^{T}C^{-1}D \right)
\end{equation} where {\it D} is the data set and {\it C} is the covariance matrix,
which represents the model of the CMB sky modulated by our observing
strategy. The covariance matrix is composed of two terms, $ C = S + N $
where {\it S} is the signal and {\it N} is the noise correlation matrix. The signal
is the convolution of the two~point correlation function and the auto~--~correlation function of the primary beam function of the interferometer.

\begin{equation}
\label{eqn_signal}
S_{ij}= C^{\rm CMB}(\theta_{ij}) \otimes S^{\rm beam}(\theta_{ij})
\end{equation}Using the band~--~power approximation where $\Delta T_\ell\equiv
\sqrt{\ell(\ell+1)C_\ell/2\pi}$ \footnote{There is a typographical error
in Dicker \et (1999), in the denominator the $8\pi$ should be
 $2\pi$ as written here.} is assumed to be constant across the range of $\ell$ covered by the window function. 
 
\begin{equation}
\label{eqn_skycorrel}
C^{\rm CMB}(\theta_{ij})=\frac {1}{2} (\Delta T)^2 \sum^{\ell_{max}}_{2} \frac{(2\ell+1)}{\ell(\ell+1)} P_{\ell}(\cos(\theta_{ij}))
\end{equation}where  $\ell_{max} $ is the limit of the summation.

\begin{figure}
\begin{center}
\setlength{\unitlength}{1cm}
\begin{picture}(5,6)(0,0)
\put(-5.5,-8){\includegraphics{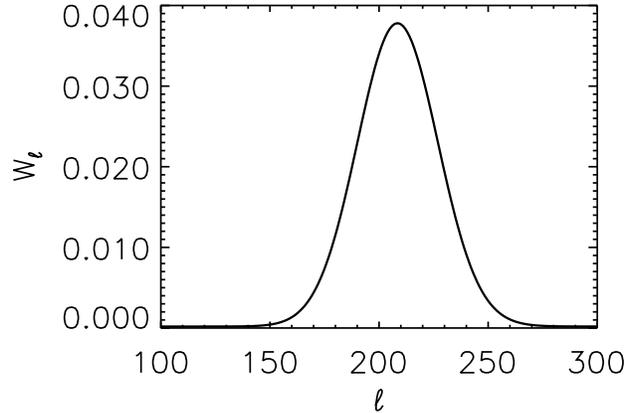}}
\end{picture}
\end{center}
\caption{The window function of the interferometer in its wide~--~spacing configuration (304\,mm $\sim 32.9$ wavelengths)which is well fit by $W_{\ell}(0)=0.037 \times {\rm exp} \left(-\frac{(\ell-208)(\ell-209)}{694}\right)$}
\label{window}
\end{figure}

The sensitivity at an individual value of $ \ell $ is given
by the window function of the interferometer in Fig.~\ref{window}. The
window function was computed using the method of Muciaccia, Natoli \&
Vittorio (1997) to decompose the beam into spherical harmonics. The resulting function can be modelled by:
\begin{equation}
\label{eqn_win}
W_{\ell}(0)=0.0377 \times {\rm exp} \left(-\frac{(\ell-208)(\ell-209)}{694}\right)
\end{equation}$W_{\ell}(0)$\, falls to half power at $\ell=208 \pm 18 $.\\

The covariance matrix can now be calculated, by numerically evaluating
equations~(\ref{eqn_signal})
and~(\ref{eqn_skycorrel}). Equation~(\ref{eqn_skycorrel}) is evaluated
to $\ell_{max}=300 $, where the sensitivity of the interferometer,
$W_{\ell}(0)$, is negligible. The result of
equation~(\ref{eqn_skycorrel}), $C^{\rm CMB}(\theta_{ij})$, is used to calculated the required value of $S_{ij}$
using equation~(\ref{eqn_signal}) . The covariance of two points with an
angular separation of $\theta_{ij}$ is then given by $ C_{ij} = S_{ij} + N_{ij}$.

\subsection{The results and the Galactic Cut}
\label{galcut}

Any analysis should take account of likely Galactic emission. At Dec +41\dg the ranges 21\hr 48\mn -- 3\hr 48\mn\, RA and 6\hr 12\mn --
19\hr 30\mn\, RA are at Galactic latitude b $\ge$ 10\dg. To find regions
free of significant Galactic emission, 48 intervals of 5 hours in RA
were analysed as described above, stepping every 0.5 hours. It was found that the above cut
was in general agreement with the higher latitude results, except the region 2\hr 48\mn
- 3\hr 48\mn\, which was excluded; this is due the Galactic plane
emission and difficulties in subtracting the contribution of 3C84, see Section~\ref{pts}.

The likelihood analysis of the ranges 21\hr 48\mn - 2\hr 48\mn\, RA and 6\hr 12\mn -
19\hr 30\mn\, RA gives $ \Delta T~=~78.5^{+12.5}_{-12.0}\, {\rm \mu} K $ and $ \Delta
T~=~69.5^{+12.5}_{-12.0}\, {\rm \mu} K $ for the cosine channel and the
sine channel respectively. The likelihood analysis using both channels
simultaneously gives $ \Delta T = 70.0^{+7.0}_{-6.5}\, {\rm \mu} K $ at
the 68 percent confidence level.

\section{The effect of foregrounds on the results}
\label{fore}

\subsection{Point Sources}
\label{pts}

The 5 strongest sources with {\it S}(33 GHz) $\geq $ 2 Jy within a 4\dg
strip centred on Dec +41\dg, listed in table~\ref{sources}, are
routinely monitored by the University of Michigan at 4.8, 8.0 and 14.5
GHz and in the Metsahovi programme at 22.0 and 37.0 GHz.  Using these
data over the period of our observations, it was possible to assess
their mean flux densities at 33 GHz. These were then convolved with the two-dimensional interferometer beam pattern centred on Dec +41\dg  and converted to antenna temperatures using the factor $ 6.90 \, {\rm \mu K Jy^{-1}} $; in this form these sources may be subtracted from the data.\\

\begin{table}
\caption{Sources within a 4$^\circ$-wide Dec strip centred on +41\degr.}
\label{sources}
\begin{tabular}{@{}l|ccc}
 Name & RA$_{\tiny 2000}$ & Dec$_{\tiny 2000}$ & Mean Flux (Jy) \\
\hline
     3C 84  & 03\hr\  19\mn\  48\rs & +41\degr\ 30\arcmin\ 42\arcsec & 11.8\\
    DA 193  & 05\hr\  55\mn\  31\rs & +39\degr\ 48\arcmin\ 49\arcsec &  -\\
   4C 39.25 & 09\hr\  27\mn\  03\rs & +39\degr\ 02\arcmin\ 21\arcsec & 9.5\\
   3C 345   & 16\hr\  42\mn\  59\rs & +39\degr\ 48\arcmin\ 37\arcsec & 9.0\\
   BL Lac   & 22\hr\  02\mn\  43\rs & +42\degr\ 16\arcmin\ 40\arcsec & 3.7\\
\hline
\end{tabular}
\end{table}

\begin{figure}
\begin{center}
\setlength{\unitlength}{1cm}
\begin{picture}(5,6)(0,0)
\put(-6,-9){\includegraphics{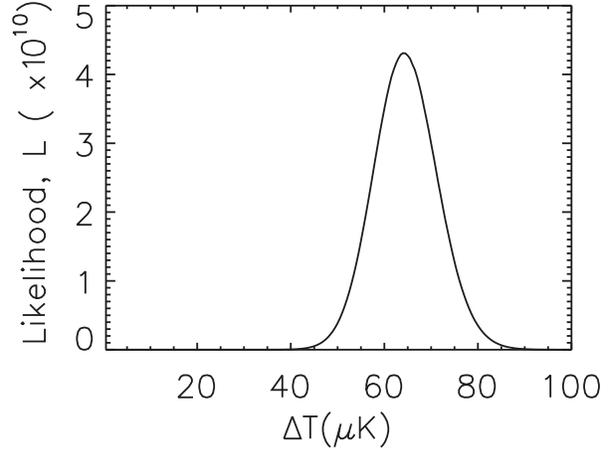}}
\end{picture}
\end{center}
\caption{The likelihood curve of the joint analysis of both channels of
the point source-subtracted data. $ \Delta T = 64.0^{+7.0}_{-6.0}\, {\rm
\mu K} $ at the 68 per cent confidence level and $ \Delta T =
64.0^{+14.5}_{-13.0}\, {\rm \mu K} $ at the 95 per cent confidence level.} 
\label{like_curve}
\end{figure}

The data ranges 21\hr 48\mn -- 2\hr 48\mn\, RA and 6\hr 12\mn -- 19\hr
30\mn\, RA were analyzed together, subtracting the point sources as
discussed above. Each channel was analyzed independently and then
combined for a joint analysis. For the cosine channel $ \Delta T =
69.5^{+12.5}_{-11.5}\, {\rm \mu K} $, for the sine channel $ \Delta T =
62.5^{+13.0}_{-11.5}\, {\rm \mu K} $ and combining both channels gives $
\Delta T = 64.0^{+7.0}_{-6.0}\, {\rm \mu K} $ at the 68 per cent
confidence level and $ \Delta T = 64.0^{+14.5}_{-13.0}\, {\rm \mu K} $
at the 95 per cent confidence level. The likehood curve of
this anaylsis is shown in Figure~\ref{like_curve}.\\

The contribution of unresolved point sources was estimated according to
the results of Franceschini et al. (1989), at 33 GHz resolution of
0.8\dg this is expected to be $ \sim 11 {\rm \mu K} $, which adds in
quadrature to the CMB signal. The contribution of unresolved sources
then accounts for approximately 1 \uk of the total signal and
accordingly the best estimate of the intrinsic CMB fluctuation amplitude
is  $ \Delta T = 63.0^{+7.0}_{-6.0}\, {\rm \mu K} $ at $\ell = 208$.

\begin{figure*}
\begin{center}
\setlength{\unitlength}{1cm}
\begin{picture}(6,12)(0,0)
\put(-5,-5){\includegraphics{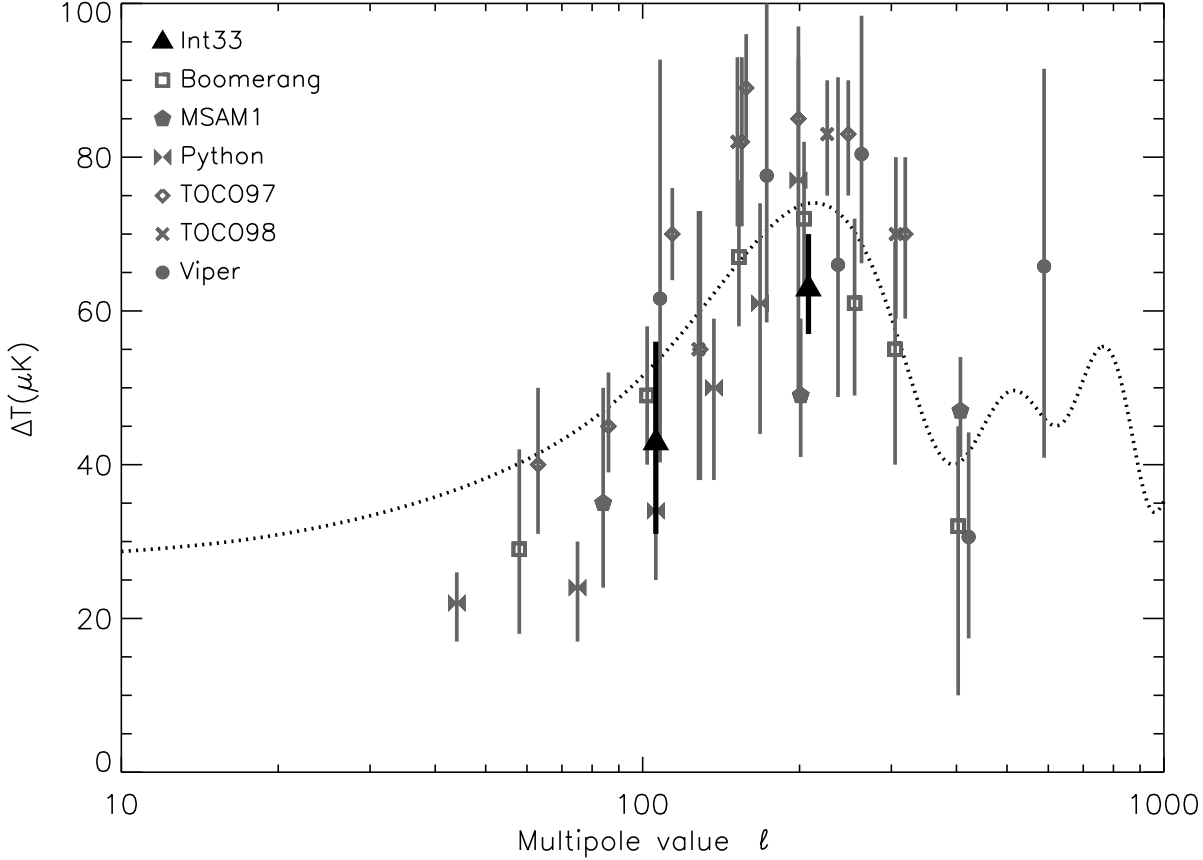}}
\end{picture}
\end{center}
\caption{Values of \DT published in the past year, as a function of
$\ell$. Our present result of $\Delta T_\ell=63^{+7}_{-6}{\rm\,\mu K}$
is shown by the heavy line at $\ell \sim 200$. The result of Dicker \et
(1999) is shown by the heavy line at $\ell \sim 100$. The dotted line represents
the model given by $\Omega_{\rm b}=0.05$,\, $\Omega_{\rm
CDM}=0.40$,\, $\Omega_{\lambda}=0.55$\, and
$H_{0}=70$\,km\,s$^{-1}\,$Mpc$^{-1}$; this is shown for illustrative
purposes and does not represent a fit to the data.} 
\label{clplot}
\end{figure*}

\begin{table*}
\caption{Values of \DT around $ \ell \sim 200 $ published in the past year.}
\label{cmba}
\begin{tabular}{@{}l|rrrl}
       Experiment & Freq.\,(GHz)  & \DT\,($\mu$K) & Multipole, $\ell$ & Reference\\
\hline
{\bf Int33}& {\bf 33} & ${\bf 63^{+7}_{-6}}$ & ${\bf 208^{+18}_{-18}}$ &
	This Paper\\
MSAM1& 156 & $49^{+10}_{-8}$ & $201^{+70}_{-82}$ &Wilson et al. (2000)\\ 
Viper&  40  & $66^{+24.4}_{-17.2}$ &  $237^{+111}_{-99}$ &Peterson et
	al. (2000)\\
Boomerang& 150 & $72^{+10}_{-10}$ &  $204^{+28}_{-21}$ &Mauskopf et al. (1999)\\
Python V& 41 & $77^{+20}_{-28}$ & $199^{+15}_{-15}$ &Coble et al. (1999)\\
TOCO98& 144 & $83^{+7}_{-8}$ & $226 ^{+56}_{-37}$ &Miller et al. (1999)\\ 
TOCO97&  35 & $85^{+8}_{-8}$ & $199^{+29}_{-38}$ &Torbet et al. (1999)\\

\hline
\end{tabular}
\end{table*}

\subsection{Spinning Dust}
\label{spin}

de Oliveira-Costa \et (1998) estimated the contribution of spinning dust
in a 19~GHz map of 3\dg resolution by correlating it with the DIRBE
sky maps, finding $\Delta T \sim 66 \pm 22  {\rm \mu K}$. Using the IRAS 100 $ {\rm \mu K}$ map, Gautier \et (1992) investigated the spatial index of the dust, finding on scales between 8\dg and 4\arcmin\, that $ \Delta T \propto \ell^{-3/2} $. This, combined with the expected spectral index of the spinning dust of $ -3.3 < \beta_{{\rm spin}} < -4$ (de Oliveira-Costa \et 1998); allowed the estimation of the contribution of spinning dust at 33 GHz and 0.8\dg resolution. The expected signal in our data due to spinning dust was found to be $\Delta T^{{\rm dust}} \approx 1.5 \pm 0.5 {\rm \mu K} $. This again adds in quadrature to the total signal, therefore the contribution from spinning dust is expected to be negligible.

\subsection{Diffuse Galactic Emission}
\label{dge}

An estimate of the amplitude of the diffuse Galactic component in our
data can be computed using the results obtained in the same region of
the sky by the Tenerife CMB experiments (Guti\'errez \et 2000). At
10.4\,GHz and on angular scales centred on $\ell = 20$ the maximum
Galactic component was estimated to be $\le 28\,{\rm\mu K}$. Assuming
that this contribution is entirely due to free-free emission $ ( \beta =
-2.1 ) $ and a conservative Galactic spatial power spectrum of
$\ell^{-2.5}$,  the predicted maximum Galactic contamination in the data
presented here is $\ 0.8\,{\rm\mu K}$, less than 2 per cent of our
measured value. Any such contribution would add in quadrature to that
from the CMB, and so is insignificant. The true make-up of the
10.4~-~GHz Galactic foreground emission will have a steeper average
spectral index since synchrotron radiation with $\beta \sim -3$\, will contribute to the measured value, therefore the contribution to our result will be even lower than stated. 
 
\section{Conclusions}
\label{conclusions}

In this paper we describe the results taken with the Jodrell
Bank~--~IAC~33~GHz interferometer in its wide spacing configuration,
which is sensitive to structure at $\ell = 208 \pm 18$. In the final
result of $ \Delta T = 63.0^{+7.0}_{-6.0}\,{\rm \mu K} $ possible
foreground contributors have been considered, the most significant of
which are point sources. In Section~\ref{pts} the contribution of
strong, monitored sources are removed from the data. To allow for the
contribution of unresolved sources the lower limit on \DT has been
increased. In Sections~\ref{spin}~and~\ref{dge}, the contributions from
dust and diffuse Galactic emission are found to be negligible. We
believe our result represents the intrinsic CMB fluctuation signal at
$\ell = 208 \pm 18 $. The quoted error in our result is dominated by
sample variance, resulting from the finite number of beam areas observed.\\  

Table~\ref{cmba} and Fig.~\ref{clplot} show our result alongside others
published in the last year from experiments covering similar angular
scales. These experiments have been made at a range of different
frequencies and regions of the sky. Our result is in good agreement with
the published data around $\ell \sim 200$. The results at $\ell \sim
200$ appear to be converging on a value of $\Delta T = 60-70 \,{\rm \mu
K}$. However, over the wider-$\ell$ range of 50~--~300 discrepancies appear
in the data sets significantly greater than the quoted errors. There
appears to be evidence for unknown systematic effects and possible
foreground contamination remaining in the data sets.

Our interferometer results show a rise in the amplitude of the power
spectrum between $\ell \sim 100 $ and $\ell \sim 200$. This is intrinsic
to the CMB since the parameters of the interferometer system,
particularly the calibration, remain the same except for the
spacing. The data in Fig.~\ref{clplot}, despite the discrepancies
referred to above, are strongly indicative of a peak in the power
spectrum at  $\ell \sim 200$. 

The interferometer is currently in its narrow spacing configuration
($\ell \sim 110 $), observing declinations spaced by 1\fdg2 from Dec
+37.4 to +43.4; these data will significantly reduce the sample variance
of the result published by Dicker \et (1999) to the order of 5 per cent.

\section{Acknowledgements}

This work has been supported by the European Community Science program
contract SCI-ST920830, the Human Capital and Mobility contract
CHRXCT920079 and the UK Particle Physics and Astronomy Research
Council. This work has been partially supported by the Spanish DGES projects
PB95-1132-C02-01 and PB98-0531-C02-02. DLH and JFMP acknowledge the
receipt of a PPARC Postgraduate Studentship. We thank Dr.H.Ter\"asranta
for providing data on point sources at 22 and 37\,GHz. This research has
made use of data from the University of Michigan Radio Astronomy
Observatory which is supported by funds from the University of Michigan.

{\bf Note added in proof}\\
Our result of $ \Delta T = 63.0^{+7.0}_{-6.0}\,{\rm \mu K}$ at $\ell =
208$ is competitive with the latest published results from the Boomerang
experiment, which found an amplitude for the first peak of $ \Delta T =
69 \pm 8\,{\rm \mu K} $ at $\ell = 197 \pm 6 $, \cite{bern00} and the
Maxima experiment, which found a peak at $\ell \approx 220$ of amplitude $ \Delta T = 78 \pm 6\,{\rm \mu K} $, \cite{hanany00}.

\end{document}